\documentclass[journal]{IEEEtai}

\usepackage[colorlinks,urlcolor=blue,linkcolor=blue,citecolor=blue]{hyperref}

\usepackage{color,array}
\usepackage{anyfontsize} 
\usepackage{graphicx}
 \usepackage{multirow} 
\usepackage{amsmath}

\begin{document}

\title{Deep Learning-based Dual Watermarking for Image Copyright Protection and  Authentication}

\author{Sudev Kumar Padhi, Archana Tiwari, and  Sk. Subidh Ali

\thanks{S. K. Padhi, A. Tiwari and S. S. Ali are with the MIST Lab, Indian Institute of Technology Bhilai, Durg 491001, India (e-mails: sudevp@iitbhilai.ac.in, archanat@iitbhilai.ac.in and subidh@iitbhilai.ac.in).}}

\markboth{Journal of IEEE Transactions on Artificial Intelligence}
%{Padhi \MakeLowercase{\textit{et al.}}: Deep Learning-based Dual Watermarking for Image Copyright Protection and  Authentication}
{ Deep Learning-based Dual Watermarking for Image Copyright Protection and  Authentication}
\maketitle
\begin{abstract}
Advancements in digital technologies make it easy to modify the content of digital images. Hence, ensuring digital images' integrity and authenticity is necessary to protect them against various attacks that manipulate them. We present a Deep Learning (DL) based dual invisible watermarking technique for performing source authentication, content authentication, and protecting digital content copyright of images sent over the internet. Beyond securing images, the proposed technique demonstrates robustness to content-preserving image manipulations. It is also impossible to imitate or overwrite watermarks because the cryptographic hash of the image and the dominant features of the image in the form of perceptual hash are used as watermarks. We highlighted the need for source authentication to safeguard image integrity and authenticity, along with identifying similar content for copyright protection. After exhaustive testing, we obtained a high peak signal-to-noise ratio (PSNR) and structural similarity index measure (SSIM), which implies there is a minute change in the original image after embedding our watermarks. Our trained model achieves high watermark extraction accuracy and to the best of our knowledge, this is the first deep learning-based dual watermarking technique proposed in the literature. 
\end{abstract}

\begin{IEEEImpStatement}

Watermarking is widely used for authentication and copyright protection. Lately, deep learning has been leveraged to perform watermarking, which yields a higher accuracy. However, existing deep learning techniques can only perform either authentication or copyright protection and are vulnerable to overwriting and surrogate model attacks.  In this work, we used deep learning to perform image copyright protection and authentication simultaneously while achieving a high level of accuracy, performance and robustness against content-preserving image manipulation, overwriting attack and surrogate model attack.  The watermarked images are indistinguishable from the original image, with an average peak signal-to-noise ratio (PSNR) of 46.87 dB and structural similarity index measure (SSIM) of 0.94 while maintaining a high accuracy of 97\% for copyright protection and 95\% accuracy for authenticating images.

\end{IEEEImpStatement}

\begin{IEEEkeywords}
Watermarking, Deep Learning, Image Authentication, Cryptographic Hash, Perceptual Hash,  Copyright Protection.
\end{IEEEkeywords}

%citation from IEEE transaction of multimedia
\section{Introduction}
\IEEEPARstart{T}{he} rapid advancement of digital technology in the domain of image processing has paved the way for creating high-resolution digital content. Side-by-side, this advanced technology also helps the attacker to launch attacks on digital content, such as illegally manipulating, duplicating, or accessing information without the owners' knowledge, violating the integrity, authenticity, and copyright protection of the target digital content. Therefore, developing secure copyright protection and authentication has become a challenging task. 

%Without secure copyright protection and authentication, verifying digital content is a major concern for law enforcement agencies and the judiciary to avoid making incorrect judgements.  During disputes over digital content, the judiciary has to validate the ownership of the content, thus protecting it's copyright.Furthermore, before making any judgement based on the digital content, the judiciary has to verify if the content is authentic, {\em i.e.,} to verify if the content is being tampered with (forged image~\cite{forgery}) or being manipulated (fake image~\cite{deepfake}). Side-by-side, it also has to validate if the content is from a genuine source. For example, while using Closed-Circuit Television ($CCTV$) footage as evidence for any incident, the judiciary also has to authenticate that the source of the footage is actually from the desired $CCTV$ camera. Therefore, before making any judgements related to digital content, the judiciary has to ensure it's copyright protection as well as it's content and source authentication.

 Digital watermarking is an effective technique for achieving copyright protection and authentication. The process of digital watermarking involves two primary operations: watermark embedding and watermark extraction. The sender embeds the watermark into the image to be protected (cover image) and sends the watermarked image to the receiver.  The original watermark is provided to the receiver or verifier in advance to perform digital content authentication or copyright protection. In order to validate the authenticity or copyright, the verifier has to extract the watermark from the received watermarked image and compare it to the original watermark. 
 
  In general, there are two types of digital watermarks addressed in the existing literature: visible and invisible watermarks~\cite{mahto2021survey,cox2002digital}. A visible watermark typically contains a visible message or a logo as a watermark in digital content. On the other hand, in invisible watermarking, a watermark is embedded and remains invisible, such that the digital media appears visually similar to the original image. Invisible watermarks can be broadly classified into fragile and robust watermarks. Fragile watermarking is used in applications where even a single-bit change in the content is not allowed~\cite{4303093,raj2021survey,Wong1998APK,ping}, as the single-bit change in the digital content will alter the watermark and will make it invalid. Therefore, fragile watermarking is used in digital content authentication and integrity verification. Robust watermarking shows robustness against content-preserving image manipulations, such as resizing, rotation, flipping, format conversion, etc~\cite{wang2002wavelet,robust,robust1,951542}, which makes it the most suitable watermarking technique for copyright protection.  However, the degree of robustness is application-specific. 
  
  Dual watermarking is another type of watermarking technique where two watermarks are embedded in the cover image to accomplish two different tasks~\cite{one-dual,sec-dual}. The most prevalent dual watermarking techniques combine a robust watermarking technique  with a fragile watermarking technique~\cite{liu2016blind,dual1,kumar2020dual,9214433,dual-2,dual-3,dual-4,dual-5,dual-6}.
 Traditional image watermarking techniques are specific to their applications and not easily adaptable to other applications. For example, the watermarking used in medical imaging can not be directly applicable to images used in remote-sensing applications. In the case of invisible watermarking, its adaptability is further aggravated as it hides the watermark at the cost of sacrificing the image quality. 
Recently, deep learning technique has evolved significantly and has wide applications. The image watermarking community has also started using it as a viable tool for watermarking as it not only provides significant improvement in performance and efficiency in comparison to traditional watermarking but is also adaptable to different applications~\cite{ref4,deepwatermark1,fang2020deep,li2021survey}. However, it also has the following challenges:
\begin{enumerate}
    \item {\em Difficulty in source authentication}: Determining whether the watermarked image is generated from the desired source~\cite{chen2023high} is challenging.
    \item {\em Vulnerable to overwriting attack:} The watermark embedded by a $DNN$ can also be overwritten using another $DNN$ to claim ownership of the image (violating copyright protection)~\cite{chen2023high, overwrite}.
    \item {\em High overhead}: It incurs high communication overhead in the form of input watermarks, which should be present with the receiver for verification~\cite{darvish2019deepsigns,wu2022attacks}.
\end{enumerate}

 We propose a deep learning-based dual watermarking technique to overcome the above issues. Our contributions to the proposed work include the following:

\begin{itemize}
\item We propose a new dual watermarking-based image authentication technique to verify content, source authentication, and protect digital content copyright.  
\item We combine the benefits of hashing and watermarking techniques. Instead of using a random watermark, we used the image's cryptographic and perceptual hash as watermarks, which eliminates the need for the original watermark and facilitates auto-verification. 
\item In our dual watermarking technique, the two extraction phases are independent of each other. The first extraction phase protects digital content copyright, while the second is used for content and source authentication.

\item Our technique is robust against content-preserving image manipulations such as noise additions, scaling, rotation, filtering, compression, change in brightness, and contrast and also secure against surrogate model attacks and overwriting attacks.
\item We demonstrated the efficacy of the proposed technique through extensive experimental results. 

\end{itemize}

The rest of the paper is organized as follows. In Section-\ref{Background}, the background related to our technique is explained. Section-\ref{motivation} gives the motivation behind the proposed work,  followed by the proposed technique in Section-\ref{Proposed Methodology}. The experimental setup and model training are presented in Section-\ref{Experimental Setup}.  The results and efficacy of our technique are discussed in Section-\ref{Results and Discussions}. We also discussed a comparative study of our work with some existing techniques in Section-\ref{comp}. Finally, the conclusions and future scope are in Section-\ref{Conclusion and Future Work}.

\section{Background }
\label{Background}
\subsection{Perceptual Hash}
\label{Hash} %updated by sir read
%Hashing is a well-known concept in the domain of cryptography~\cite{lin1998generating}, through which a unique digest of a given message is generated. Cryptographic hash functions are generally used in digital signatures, message authentication codes, and to check the integrity of a message, image, or digital file~\cite{crypto}.  
Secure Hash Algorithm ($SHA$) is one of the standard cryptographic hash functions with a strong avalanche effect, which means any tiny change to the input will lead to a significant change at the output (hash value)~\cite{chen2022asymmetric,crypto,lin1998generating}. When it comes to image processing, cryptographic hash becomes less useful, specifically where certain perceptual content-preserving image manipulations, such as JPEG compression, noise, blurring, cropping, deformations, etc, are allowed as these manipulations will alter the hash value. In this line, the perceptual hash is a class of comparable hash functions that generate a distinct and unique hash using robust or invariant features present in the input image~\cite{phash2, thesis,phash1,phashcopy2,ref8,qamra2005enhanced} such that two visually similar images should produce the same perceptual hash value. There are four basic techniques for generating perceptual hash: frequency transformations, dimensionality reduction, statistical methods,  and extracting the local features of an image~\cite{ref10}.

The most prominent technique for calculating the perceptual hash is the Discrete Cosine Transformation ($DCT$) technique~\cite{phash-dct,phash-dct1,phash-dct3}. In this technique, the input image is expressed in terms of cosine functions oscillating at different frequencies. Frequency coefficients representing the lower frequency components contain important features of an image that are robust against distortions. In $ DCT$-based perceptual hash, the input color image is first converted into a grayscale image and resized to $32 \times 32$ pixels. The grayscale image is then converted to the $32 \times 32$ frequency domain image through $DCT$ transformation. The   $8 \times 8$ block of pixels in the upper-left portion of the image represents the low-frequency block of the resized grayscale image. Each frequency coefficient present in the low-frequency block of pixels is compared to the mean value of all the frequency coefficients of pixels in the low-frequency block.  If the value of an individual frequency coefficient is greater than the mean value, its phase is taken as $1$, and if it is not, it is taken as $0$. The final hash is obtained by arranging the 64-bit binary values sequentially, thus representing the perceptual hash of an image.

%The obtained perceptual hash will not vary if the image remains perceptually the same.

%\subsection{ Dual Image Watermarking}
%Dual watermarking approaches started with embedding a visible and an invisible watermark into the cover image~\cite{one-dual} followed by embedding one watermark in the spatial domain and the other in the frequency domains~\cite{sec-dual}. 
%The most popular dual watermarking techniques involve a combination of a robust watermark and a fragile watermark~\cite{liu2016blind,dual1,kumar2020dual,9214433,dual-2,dual-3,dual-4,dual-5,dual-6}. There are many techniques to perform dual watermarking. One of the approaches is to combine both watermarks into a single watermark and embed it in the image. However, in this technique, the two watermarks are combined into one, and hence, attacking the embedded watermark will affect both watermarks. Another approach is to embed the two watermarks in different regions of the image simultaneously to prevent overlapping. Both of these techniques have their advantages and can be chosen based on specific requirements and objectives. The most popular technique is to embed the first watermark and obtain a watermarked image, which is then used as a cover image to embed the second watermark to produce the final watermarked image. In this technique, the extraction scheme for watermarks largely depends on the way in which watermarks are embedded within the image and makes the watermark embedding more dynamic. 

\subsection{Deep learning-based watermarking}

\begin{figure}[htb]
\centering
\includegraphics[scale=0.19]{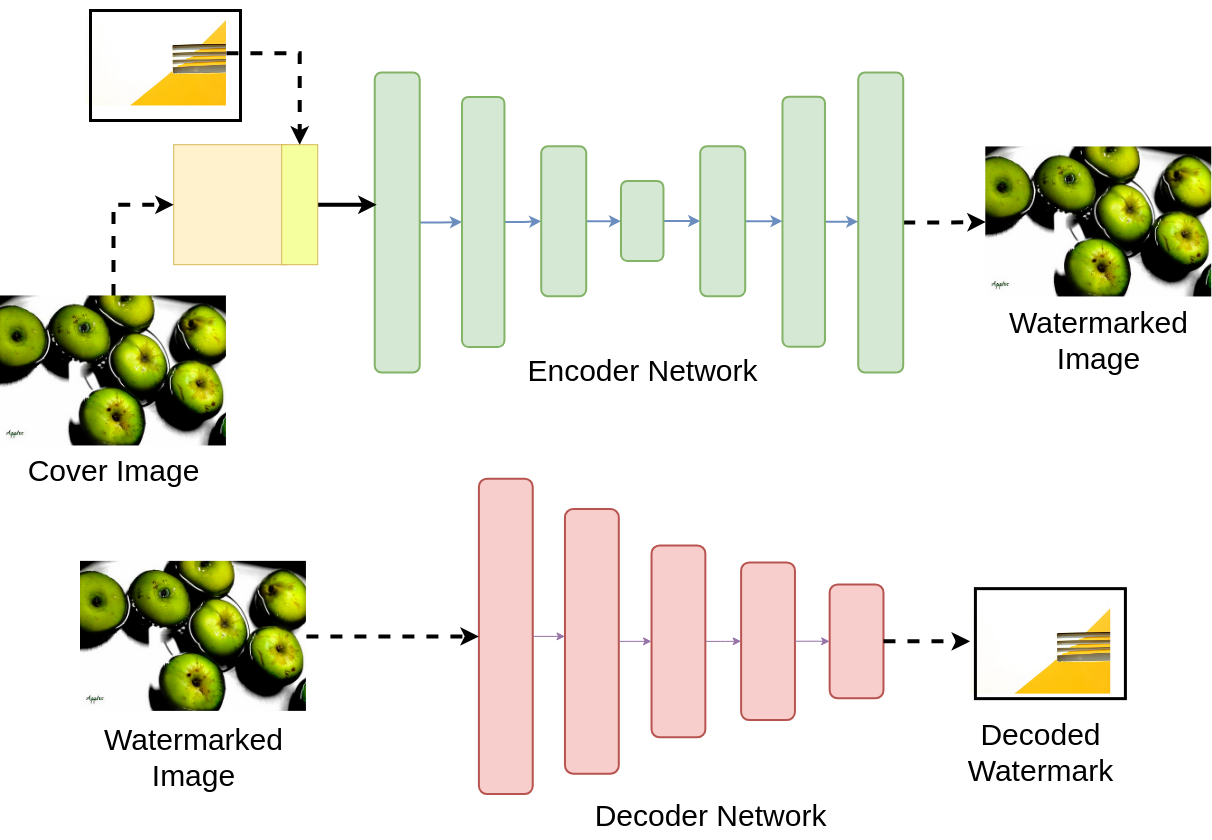}

\caption{General overview of deep learning-based watermarking where an encoder network is used for embedding the watermark and the decoder network is used for extracting the watermark.} %centre caption above image
\label{nn2}
\end{figure}

 Deep learning-based watermarking techniques are the newest approach that can efficiently watermark images by overcoming the drawbacks of traditional watermarking techniques~\cite{ref4,deepwatermark1,fang2020deep,li2021survey}. In deep learning-based watermarking, encoder and decoder networks are used to perform watermarking, as shown in Fig-\ref{nn2}. The encoder network constructs a watermarked image by embedding the watermark into the cover image, as shown in the upper part of Fig-\ref{nn2}.  Once the watermarked image is constructed, the decoder network (shown at the bottom part of Fig-\ref{nn2}) is used to extract the watermark from the watermarked image.
Deep learning-based watermarking has its own challenges. The fragile nature of $DNN$  can sometimes cause the model to fail~\cite{lacuna2}. As a result, fully utilizing the ability of deep neural networks to learn automatically and generalizing it to both watermark embedding and extracting processes is difficult. The lack of a good balance of imperceptibility and robustness is also a major concern in deep learning-based watermarking~\cite{lacuna1,lacuna6,deepwatermark1,lacuna4,lacuna5}. 

\section {Motivation}
\label{motivation}
Traditional image watermarking methods face challenges due to their limited robustness against content-preserving image manipulation. For example, extraction can only tolerate certain types of attacks; thus, rotation to certain angles or blurring of the image to a larger degree may result in distortions in embedded watermarks.  Deep learning-based watermarking is more robust to content-preserving image manipulations~\cite{lacuna1,lacuna2,lacuna6,lacuna4},  
however, maintaining a balance between robustness and imperceptibility is a challenge~\cite{deepwatermark1}. Although the introduction of deep learning in watermarking has significantly improved the security of watermarking techniques,  it lacks inherent watermark verification. Therefore, the verifier has to have the watermark with it; only then it can verify the watermark by manually extracting the embedded watermark from the watermark image and comparing it with the watermark in hand. %Sending the watermark to the verifier will also increase the communication overhead. Moreover, there is no relation between the watermark and the cover image; therefore, any random image can be used as a watermark. This can increase the chance of misunderstanding between the sender and receiver if a new watermark is used every time. 

%In the same line, while using deep learning-based watermarking for protecting digital content copyright, 
Overwriting attacks have also shown vulnerabilities in watermarking~\cite {chen2023high,overwrite}, where the attacker trains a surrogate encoder to overwrite the original watermark with its own watermark, such that the target decoder extracts the attackers watermark instead of the original watermark. Therefore, the attacker can make a false claim on the watermarked image.  Additionally, deep learning-based watermarking techniques are at risk of surrogate model attacks~\cite{chen2023high}, in which the decoder fails to differentiate between the images generated from the original encoder and the surrogate encoder, which leads to successful extraction of the watermark generated from the surrogate encoder. Robust training makes the watermark easy to extract, which can lead to a surrogate attack as the decoder is robust and always extracts the watermark. This vulnerability highlights the need for source authentication to ensure the authenticity of the source generating the watermarked image. 

%Using random images as watermarks introduces additional vulnerabilities that an attacker can misuse to launch an attack. These vulnerabilities can be overcome by using a meaningful watermark where certain cover image features are used as the watermark~\cite{phash-dct,phash-dct1,phash-dct3,phash-dct2}.

Based on the above-mentioned points, it can be concluded that deep learning-based watermarking technique achieves significant improvements over traditional watermarking; however, it is not entirely secure against existing attacks. In order to improve the security and robustness of the deep learning-based watermarking technique, it should have the following key features:

\begin{enumerate}
    \item {\em Feature-based watermark:} Instead of random images,  salient features, such as edges, textures, or fractal dimensions,  should be used as watermarks. This reduces the requirement of selecting a new watermark and increases its security~\cite{meanngfulwwatermark, meanngfulwwatermark1}. 
    \item {\em Multiple watermarking:} Facilitating copyright protection along content and source authentication requires at least dual watermarking.
    \item {\em Source authentication :} Incorporating source authentication into the watermarking technique will ensure protection against surrogate model attacks.  
    \item {\em Auto verification of watermark:} The watermark verification should be performed by a unique decoder automatically at the receiver's end without requiring original watermarks. Hence, communication overhead is reduced.
    \item {\em Robustness against content-preserving image manipulation:} Deep learning-based watermarking techniques should perform watermark verification successfully even if the watermarked images are maliciously manipulated or tampered with to a certain extent.

\end{enumerate}
\section{Proposed Methodology} \label{meth}
%The method section requires improvement as it is unclear and lacks crucial details. The overall pipeline needs clarification, and the inclusion of equations can enhance the explanation of the method.
\label{Proposed Methodology} 
We propose a deep learning-based invisible dual watermarking technique that has all the features mentioned above.  This dual watermarking technique serves three objectives at once: 1) protecting digital content copyright, 2) content, and 3) source authentication. Perceptual hash is used as the first watermark for protecting digital content copyright, whereas cryptographic hash is used as the second watermark for content and source authentication. 
\begin{figure*}[!htb]
\centering
\includegraphics[scale=0.125]{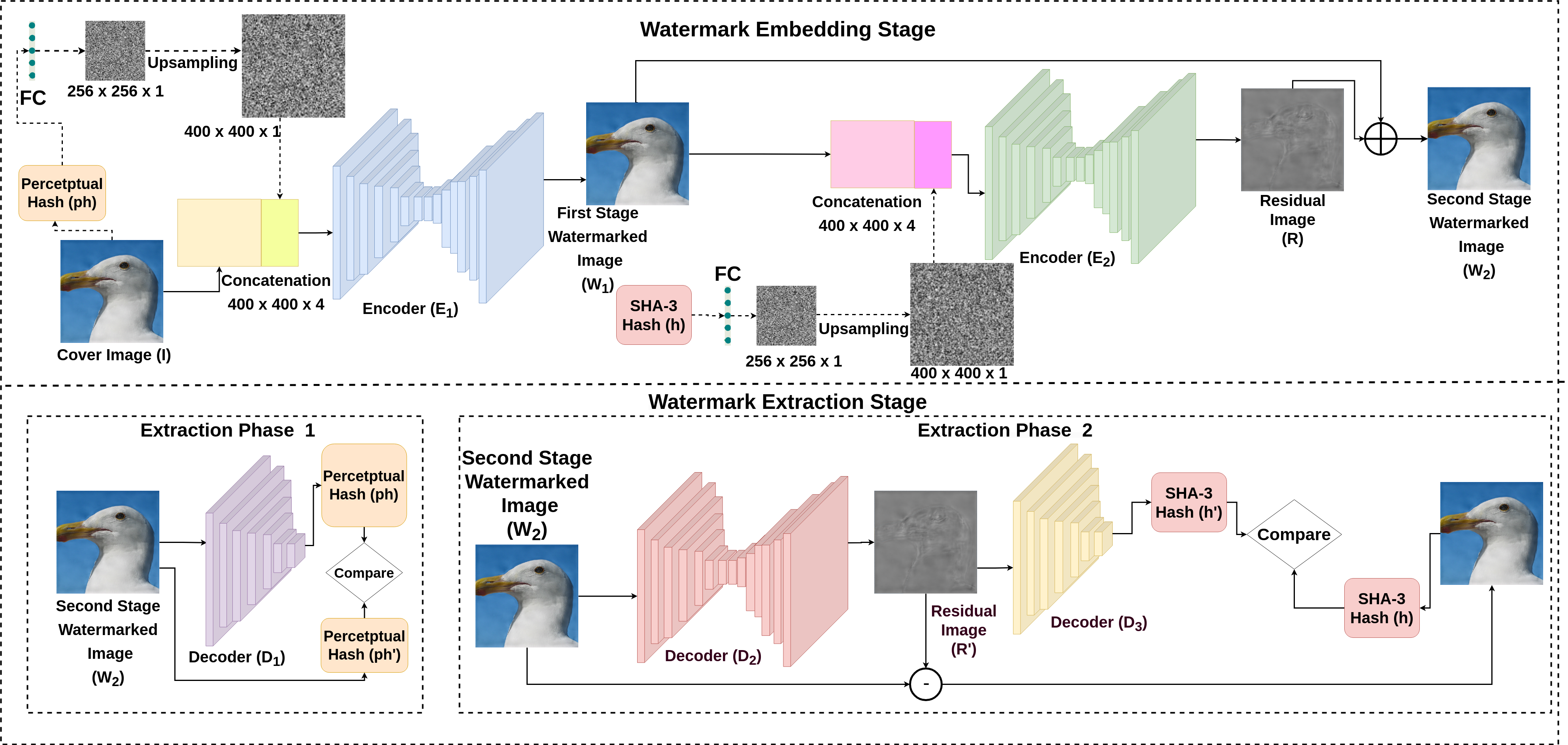}
\centering
\caption{Overview of the proposed Deep Learning-based Dual Watermarking. In the watermark embedding stage, two encoders, $E_1$ and $E_2$, are used to embed perceptual and cryptographic hash, respectively. There are two phases for the watermark extraction stage. In the first phase, decoder $D_1$ is used to verify the image ownership for copyright protection, while in the second phase, decoders $D_2$ and $D_3$ are used for content and source authentication. It is to be noted that both phases of watermark extraction are independent of each other.} %centre caption above image
\centering
\label{model}
\end{figure*}

 %In the watermark embedding stage,  a cover image ($I$) is used for embedding the watermark. The perceptual hash  ($ph$) of $I$ is used as the first watermark.  $ph$  is  embedded into $I$ by using the encoder ($E_1$). The reconstructed image obtained from $E_1$ is the first stage watermarked image ($W_1$).  The second watermark is the cryptographic hash ($h$) of $W_1$. The encoder ($E_2$) embeds $h$ into $W_1$. Hence, we obtain a  second stage watermarked image $W_2$, which will protect digital content copyright and authentication of content and source. The watermark embedding stage is explained in detail in Section \ref{emb}.
%, in the first phase, one decoder ($D_1$ ) is used. While in the second phase, two different decoders ($D_2$ and $D_3$) are used for extracting the watermark.  $D_1$ performs extraction of the perceptual hash from $W_2$, which serves the purpose of protecting digital content copyright. At the same time, $D_2$ and $D_3$ are used to extract the cryptographic hash from $W_2$ for content and source authentication. 

\subsection{Watermark Embedding} \label{emb}
The watermark embedding process is shown in the upper part Fig-\ref{model}. The first step in our technique is to find the perceptual hash of the cover image $I$, which is our first watermark. There are different techniques for calculating the perceptual hash of an image.  We have chosen $DCT$-based perceptual hash as discussed in Section-\ref{Hash}. It has advantages in terms of robustness against distortions and attacks caused by content-preserving image manipulation. We compute $64$-bit $DCT$-based perceptual hash $ph$  of $I$, which is passed into a pre-trained Fully Connected layer ($FC$) to produce $256 \times 256 \times 1$  tensor and then upsampled to a $400 \times 400 \times 1$ tensor. The upsampled tensor is concatenated to $I$ as the fourth channel and passed to the encoder $E_1$. The job of $E_1$ is to embed $ph$  into $I$ as a watermark and produce a $400 \times 400 \times 3$ image as output, which is our first stage watermarked image $W_1$. It is ensured that $W_1$ will have the same perceptual hash as  $I$ by using three loss functions, which are residual regularization,  LPIPS perceptual loss, and critic loss. More details regarding these loss functions are discussed in Section-\ref{arch}. 

Next, we compute the $256$-bit cryptographic hash $h$ of $W_1$ using $SHA$-$3$ and use it as our second watermark. The  $h$ is passed through the same pre-trained $FC$ to produce a $256 \times 256 \times 1$  tensor that is upsampled to  $400 \times 400 \times 1$ tensor. The upsampled tensor is concatenated to  $W_1$  as the fourth channel and  $400 \times 400 \times 4$ tensor is given as input to the second encoder $E_2$, which produces a residual image $R$ of resolution $400\times400\times3$ as the output.  The obtained $R$ is added to $W_1$, which yields the second stage watermarked image $W_2$ that is perceptually similar to the $I$. This addition of residual image ($R$)  does not change the perceptual hash and perceptual similarity of  $W_2$  with respect to $I$.  This is ensured by using appropriate loss functions while training. Adding to that,  we have also used two different techniques to embed the watermarks, thus reducing the chance of overlapping i.e, one watermark distorting the other. The technique followed for embedding $h$ into $W_1$ is inspired by the approach of Stegastamp~\cite{stegastamp}. Dual watermarking makes it convenient to perform different verification in one watermarked image. In our case, $W_2$ protects digital content copyright, content and source authentication.

\subsection{Watermark Extraction}
\label{auth}
%Write about some point
The watermark extraction and verification process is shown in the bottom part of Fig-\ref{model}, which consists of two independent extraction phases: 
\begin{enumerate}
    \item {\em Extraction phase $1$:} is used for digital content ownership verification. We compute the perceptual hash ($h^\prime$) of $W_2$ using DCT transformation. Then $D_1$ extracts the embedded perceptual hash ($h$) from $W_2$. If the extracted and the computed perceptual hashes of  $W_2$ match, {\em i.e.,} $ph={ph}^\prime$, the image ownership is claimed.
    \item {\em Extraction phase $2$:} is used for content and source authentication for the given dual watermarked image $W_2$. It uses two decoders: $D_2$ and $D_3$.  $D_2$  reconstructs a residual image ($R^{\prime}$) from given $W_2$.   $R^{\prime}$ is passed into $D_3$, which decodes the cryptographic hash ($h$) present in $R^{\prime}$. Subsequently, $R^{\prime}$ is subtracted from $W_2$, and the $SHA$-$3$ hash value ($h^\prime$) of the result is computed. If the cryptographic hash ($h$) extracted by  $D_3$ matches with the computed cryptographic hash ($h^\prime$), {\em i.e.,} $h={h}^\prime$, the authenticity of the content and source of the image is verified. 
\end{enumerate}

A robust copyright protection mechanism should be able to extract the watermark successfully from the watermarked image and claim ownership even if the watermarked image is tampered with. Hence, $D_1$ used in the {\em Extraction phase $1$} should successfully extract the watermark from the tampered image to protect digital content copyright. To achieve this,  we have used a DCT-based perceptual hash, which is robust against distortions, compression, white Gaussian noises, and Gaussian blurring.  Moreover, the spatial transformer provides invariance against various image transformations such as translation, scaling, rotation, cropping, change in brightness and contrast and deformations, as discussed in the next Section-\ref{arch}. To verify the authenticity of the content and source for an image, the hash of the image should not change. Therefore $D_2$ and $D_3$ are used in the {\em Extraction phase $2$}. It is ensured that  $D_2$ fails to reconstruct the residual image if the image is tampered with. Thus both $D_2$ and $D_3$ are trained on the images without adding any distortions.

\section{Experimental Setup}
%The experiments section is not comprehensive. While it is acknowledged that existing methods based on invertible networks may not be directly applicable to dual watermarking, the authors should make an effort to reproduce some of these methods and provide comparisons to demonstrate the effectiveness of their approach.
\label{Experimental Setup}
In this section, we present the experimental results of our technique.  The model is trained in Python $3.6$ with Pytorch~\cite{pytorch}. Experiments are conducted on a machine with two $14$-core Intel i$9-10940X$ CPUs, $128$ GB RAM, and two $Nvidia$ $RTX$-$5000$ GPUs with $16$ GB VRAM each. 

\subsection {Dataset} \label{data}

 To explore the generality of our technique, we performed our evaluations on two datasets, CelebA~\cite{celeba} and Mirflickr~\cite{mir}. CelebA contains $200$k  face images of various celebrities with different poses and backgrounds at a resolution of $178\times218$. Facial images are the ones that are mostly tampered and identifying the original source image will be a great help in discarding the fabricated media. During training, the images are upsampled to $400\times400$, which shows that our technique can also be used for embedding dual watermarks in low-resolution images without creating distortions. The Mirflickr dataset consists of one million images from the social photography site Flickr, from which we have chosen $200$k random images for training and testing our technique. It consists of different types of images with varied contexts, lighting, and themes, perfect for exhibiting our techniques' generalization. The images are of resolution $500\times500$, which is downsampled to $400\times400$ for training.
 
 We have three instances of our proposed deep learning-based dual watermarking technique. The first instance of the model is trained on {\em CelebA}, the second  on {\em Mirflickr}, and the third on the dataset consisting of images from both {\em CelebA} and {\em Mirflickr}.   While training the first two instances of the model,  $150$k images are chosen as the training set and the remaining $50$k images are chosen as the test set. For the third instance of our model,  we have taken $100$k random images each from  {\em CelebA} and {\em Mirflickr} datasets. From the combined $200$k images, $150$k images are chosen as the training set and the remaining $50$k images are chosen as the test set.

 \subsection{Network Architecture} \label{arch}
Our model consists of five networks ($E_1$, $E_2$, $D_1$, $D_2$ and $D_3$), which are trained in an end-to-end manner. We employ the U-Net~\cite{unet} architecture for both $E_1$ and $E_2$ with skip connections between the encoder and decoder layers of each U-net. It helps $E_1$ and $E_2$ to generate watermarked images that are perceptually similar to the input cover image.  Furthermore, the skip connections also preserve crucial spatial data and aid in recovering minute features that would have been lost during the downsampling of the image by the U-net encoders.

%While embedding the perceptual hash, the emphasis is on the perceptual similarity between the reconstructed and original cover image, such that the perceptual hash does not change. In the same line, the hash is embedded and the residual image $R$ is generated by $E_2$, which is added to $W_1$ to produce $W_2$. 

We employ a spatial transformer~\cite{spatial} to extract the perceptual hash, which takes $W_2$ as an input and extracts its embedded perceptual hash $ph$.   The spatial transformer used in $D_1$ provides invariance against various image transformations such as translation, scaling, rotation, cropping, deformations, change in brightness and contrast, etc. Decoding the cryptographic hash involves reconstructing $R^{\prime}$ using decoder $D_2$ and extracting $h$ from $R^{\prime}$ using $D_3$. U-Net is used to reconstruct $R^{\prime}$ using $D_2$,  whereas we use a simple convolution neural network with five convolution layers followed by two fully connected layers for extracting the hash from $R^{\prime}$ using $D_3$.

\subsection{ Loss Function and Scheme Objective} \label{loss}

As mentioned earlier,  the model is trained in an end-to-end manner.  Generally,  embedding watermarks distorts the image, however, in our case,  the perceptual hash and cryptographic hash introduce a  small distortion in the cover image ($I$) such that the perceptual hash remains the same. Embedding the first watermark requires  $I$ and the perceptual hash ($ph$) of $I$, which outputs $W_1$. To ensure minimal distortion while embedding $ph$ in $I$ as a watermark,  we used   $L_2$ residual regularization ( $L_R$), the LPIPS perceptual loss ($L_P$) and a critic loss ($L_C$) functions. $L_R$ reduces the chance of the encoder to overfit while $L_P$ and  $L_C$ increases the perceptual similarity between $I$ and $W_1$.  $L_P$  help us to compute the perceptual similarity between the two images corresponding to a predefined network. We use VGG-$19$ in our case~\cite{lpips} as the predefined network. For given two images $A$ and $B$, the $L_p$ corresponding to the $VGG$-$19$ is computed as follows: 

\begin{equation}\label{lpips}
    L_{P} = \sum_{k} \tau^k (VGG^k (A) - VGG^k (B))
\end{equation}

In Equation~\eqref{lpips},  features are extracted from $k$ layers of the $VGG$-$19$ network. The function  $\tau$ transforms deep embedding to a scalar LPIPS score. The final score between two images ($A$ and $B$) is computed and averaged for $k$ layers. {\em $L_C$} is calculated using a simple five-layer deep convolution neural network which classifies cover images and watermarked images using  Wasserstein loss~\cite{wass}.  It predicts whether a watermark is encoded in an image such that the embedding of the watermark can be improved.

\begin{equation} \label{eq1}
\begin{split}
    L_{E_1} =  L_R (E_1) + L_P(W_1, I) + L_C(I , W_1)
\end{split}
\end{equation}
Equation-\ref{eq1} refers to the loss function used for $E_1$. Similarly, embedding the second watermark requires cryptographic hash ($h$) of $W_1$ and $W_1$, which outputs $W_2$. Thus we have used the same  $L_R$, $L_P$, and $L_C$ loss for the encoder $E_2$ as referred in Equation-\ref{eq2}.
\begin{equation}    \label{eq2}
\begin{split}
    L_{E_2} =  L_R(E_2) + L_P(W_2, I) + L_C(I , W_2))
\end{split}
\end{equation}
$D_1$ uses binary cross-entropy loss to extract the embedded perceptual hash ($ph$) from $W_2$ (Equation-\ref{eq3}). Decoder $D_2$ is used for reconstructing  $R^{\prime}$ from $W_2$ while decoder $D_3$ is used for extracting the embedded cryptographic hash ($h$) from $R^{\prime}$. After reconstructing  $R^{\prime}$ using $D_2$ it is compared with the original $R$ using $L_1$ loss, {\em LPIPS} loss, and Mean Square Error ({\em MSE}) (Equation-\ref{eq4}). Then $D_3$ is used to extract $h$ from $R^{\prime}$,  thus binary cross-entropy loss is used (Equation-\ref{eq5}). 
\begin{equation} \label{eq3}
L_{D_1} =   L_{BCE}(D_1(W_2),ph) 
\end{equation}
\begin{equation} \label{eq4}
    L_{D_2} = L_1(R , R^{\prime}) +  L_P(R , R^{ \prime}) + MSE(R , R^{\prime}) 
\end{equation}
\begin{equation} \label{eq5}
     L_{D_3} = L_{BCE}(D_3(R^{\prime}),h) 
\end{equation}

Training loss(Equation-\ref{eq6}) is the  sum of all five loss components ($L_{E_1},\: L_{E_2},\: L_{D_1},\: L_{D_2}, and \: L_{D_3}$), which is minimized and $\lambda_1$, $\lambda_2$, $\lambda_3$, $\lambda_4$ ,and $\lambda_5$ are weight factors. 
\begin{equation} \label{eq6}
L=\lambda_1 L_{E_1} + \lambda_2 L_{E_2}+ \lambda_3 L_{D_1} + \lambda_4 L_{D_2} + \lambda_5 L_{D_3}
\end{equation}

 For integrity and authentication, the reconstruction of the residual image should be identical. We have used $L_1$, $MSE$ and $LPIPS$ loss for that. Adding to that, $SHA$-$256$ is not directly applied to the first stage watermarked image. Rather, a pre-trained autoencoder ($CNN$-based) is used, which obtains the continuous-valued representation of the image in the latent space. This representation is a set of real numbers that captures essential features of the image.  Then, these numbers are normalized before the quantization step to ensure a consistent mapping of each continuous value to a specific discrete value. Then, $SHA$-$256$ is applied to generate a fixed-size hash value. As $D_2$ will reconstruct identical residual images, which will have the original hash that is extracted by $D_3$.

\subsection{Training of network}\label{train}

%Our technique consists of two encoders and three decoders, which are trained together, as explained in Section \ref{meth}. To ensure that the proposed technique works, we need to minimize the loss functions such that the model converges smoothly. The training procedure starts with embedding both the perceptual and cryptographic hash as watermarks into the cover image, as explained in Section \ref{emb}. This is followed by the extracting phase, as discussed in Section \ref{auth} for protecting digital content copyright as well as content and source authentication. 

In order to protect digital content copyright, $D_1$ should extract the watermark even if $W_2$ is subject to content-preserving image manipulation such as JPEG compression, noise, blurring, cropping, deformations, etc. Hence, while training,  we performed random blurring, adding random noise, flipping, rotating, cropping, and compression to random watermarked images in the datasets such that $D_1$ can be robust in extracting the embedded perceptual hash. To reduce the impact of resizing and cropping on watermarked images,  we have restricted the embedding of perceptual hash in the centre portion of the images while training~\cite{stegastamp}. Decoders $D_2$ and $D_3$ are trained on the watermarked images without adding any distortions. This is due to the fact that minute changes in the watermarked image should be detected for successful content and source authentication.  We have trained three instances ({\em CelebA},  {\em Mirflickr} and combined $200$k images) of our model for $200$ epochs with a batch size of $32$. Early stopping and  $10$-fold cross-validation are employed to prevent the model from overfitting.  We have used Adam optimizer to optimize the parameters of all five sub-models in our technique. The learning rate is set as $1.0 \times 10^{-5}$, and we decay the learning rate by $0.2$ if the loss does not decrease within $10$ epochs. 

\subsection{Challenges Faced During Training}
 The main challenge in our proposed technique is to embed two watermarks without degrading the quality of the cover image and using different decoders to extract both watermarks independently. The performance of our model depends on the successful convergence of the five loss functions used, which will ensure high accuracy. The steps involved in embedding and extracting the perceptual hash are similar to the techniques used in generic deep learning-based watermarking~\cite{fang2020deep,deepwatermark1,li2021survey,ref4}. The challenge here is to extract the perceptual hash from the second stage watermarked image after embedding the cryptographic hash. We must ensure that the perceptual and the cryptographic hash embedding do not overlap and are independent of each other. Hence to extract the perceptual hash successfully, both watermarks are embedded using different techniques. In the first stage, the encoder reconstructs $W_1$. In contrast, for the second stage,  the encoder generates {\em R}, which is added to  $W_1$ to obtain $W_2$. This assures that the watermarks do not overlap and the model can easily extract both watermarks. The difficulty is also faced while embedding the cryptographic hash and extracting it. As we know, the cryptographic hash changes with the slightest change in the image. Thus, $D_2$  must reconstruct the $R^{\prime}$  identical to $R$ before decoding the cryptographic hash. To achieve this, we have used {\em L$1$}, {\em LPIPS}, and {\em MSE} loss as the metrics while training the model to make $R^{\prime}$ identical to {\em R}.  
%usng deep learning also introduced new problems like surrogate attack.
%challenge about robustness write precise points

\begin{table*}[t]
\small
\caption{Accuracy of our watermarking technique along with the image quality of the watermarked image.}
\centering\begin{tabular}{|l|c|c|c|c|c|c|c|}
\hline
\textbf{Dataset}                        & \textbf{\begin{tabular}[c]{@{}c@{}}Training Accuracy\\  ($D_1$)\end{tabular}} & \textbf{\begin{tabular}[c]{@{}c@{}}Test  Accuracy\\($D_1$)\end{tabular}} & \textbf{\begin{tabular}[c]{@{}c@{}}Training Accuracy\\  ($D_2$ and $D_3$)\end{tabular}} & \textbf{\begin{tabular}[c]{@{}c@{}}Test  Accuracy\\ ($D_2$ and $D_3$)\end{tabular}} & \multicolumn{1}{l|}{\textbf{SSIM}} & \multicolumn{1}{l|}{\textbf{PSNR}} & \multicolumn{1}{l|}{\textbf{MSE}} \\ \hline
\textbf{CelebA}                       & 99.3                                                                               & 97.4                                                                           & 97.3                                                                                     & 95.4                                                                                 & 0.94                               & 46.3                               & 0.07                              \\ \hline
\textbf{Mirflicker}                   & 99.1                                                                               & 96.9                                                                           & 96.6                                                                                     & 94.9                                                                                 & 0.95                               & 47.4                               & 0.01                              \\ \hline
\textbf{Merged} & 99.1                                                                               & 97.2                                                                           & 96.9                                                                                     & 95.2                                                                                 & 0.95                               & 46.9                               & 0.05                              \\ \hline
\end{tabular}

\label{table:res}
\end{table*}

\section{Results and Discussions}
\label{Results and Discussions}

%The proposed technique has two goals: protecting digital content copyright and ensuring content and source authentication of the received image. To do this, we employed dual watermarking, where the visual quality of the cover image and the watermarked image remain indistinguishable. The perceptual hash is used as the watermark for protecting digital content copyright, whereas content and source authentication are ensured by using the cryptographic hash as the watermark.

%The first goal is achieved by comparing the perceptual hash that is extracted from the watermarked image using decoder $D_1$ with the computed perceptual hash of the watermarked image. $D_1$ is robust against scaling, cropping to a certain point, rotating, filtering, noise addition, change in brightness and contrast and image modifications. Hence, the embedded perceptual hash can be extracted successfully for verification, meeting the objective of protecting content copyright. 

\begin{figure}[!htb]
\centering
\includegraphics[scale=0.36]{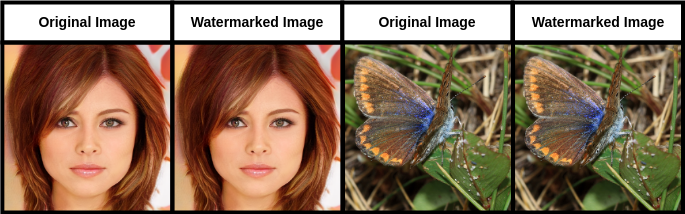}
\caption{Visual quality of the image after embedding the watermarks using our technique.}
\label{watermark}
\end{figure}

Fig-\ref{watermark} shows the visual quality of the cover image and the watermarked images after embedding the perceptual and the cryptographic hash as the watermark. To verify the received watermarked image quality, we used structural similarity index  ({\em SSIM})~\cite{ssim} and peak signal-to-noise rate ({\em PSNR}) matrices. {\em SSIM} to compare local patterns of pixel intensities, which represent the attributes that represent the structure of objects in the image, along with luminance and contrast. The higher the value of {\em SSIM}, the better and the value closer to one represents the two images are structurally the same. On the other hand, {\em PSNR}  measures the distortion between the cover and the watermarked image such that the quality of the reconstructed image can be validated. The higher the {\em PSNR} (value greater than $30$ dB is considered good~\cite{psnr-ssim}), the better the quality of the reconstructed image. Mean squared error ({\em MSE}) is used to find the change in pixel distribution. Thus, the level of distortion at the pixel level between the cover and the watermarked image is validated. The lower value of {\em MSE} represent less distortion in the reconstructed image. 

Our technique works only if the computed perceptual and cryptographic hash match bit by bit with the extracted perceptual and cryptographic hash, respectively. Hence, the accuracy of our technique means successful matching between extracted and computed hashes. Table-\ref{table:res} shows the training and testing accuracy of our model for various instances. The higher accuracy for decoders shows the success of extracting and comparing the watermark with more than $96$\% accuracy for $D_1$ and more than $94$\% for $D_2$ and $D_3$. Table-\ref{table:res} also demonstrates the similarity between the cover image and the watermarked image, which is shown quantitatively as {\em SSIM}, {\em PSNR}, and {\em MSE}. Higher {\em SSIM} and {\em PSNR} values show that the watermarked image has good visual quality, and low {\em MSE} indicates that our technique adds little distortion to the watermarked image with respect to the cover image.

\subsection{Robustness Against Content Preserving Image Manipulations}
\begin{figure}[!htb]
\centering
\includegraphics[scale=0.24]{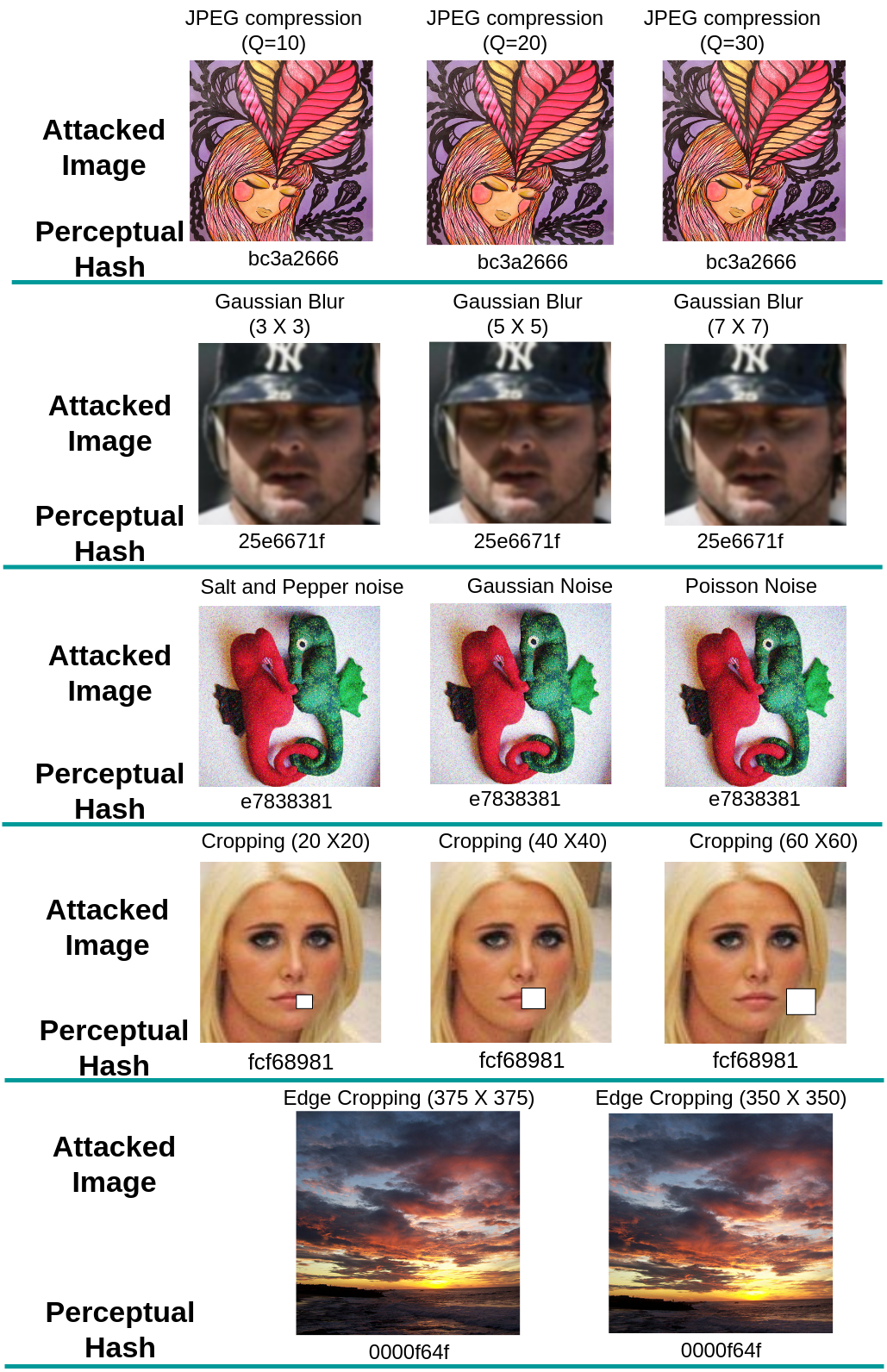}
\caption{Perceptual hash of the image when different content-preserving image manipulations are performed. }
\label{watermark_att}
\end{figure}

The most crucial issue in copyright protection systems is robustness, which signifies a watermarking system's capacity to withstand various attacks. Attackers may attempt to remove the embedded watermark by applying content-preserving image manipulations in the watermarked image.  $D_1$ is trained against various content-preserving image manipulations as shown in the first column of Table-\ref{table:rob}, which are  standard across multiple watermarking techniques~\cite{deepwatermark1,ref4, liu2016blind,kumar2020dual, 9214433}. For JPEG compression, three different image quantization factor (Q) values of $10$, $30$, and $50$ are used to control the amount of lossy compression applied to the image. A higher Q-value results in less compression and better image quality, and vice versa. Gaussian blur with window sizes $3 \times 3$, $5 \times 5$, and $7 \times 7$ are applied. Random crops of size $20 \times 20$, $40 \times 40$, and $60 \times 60$ are performed. Cropping the edges also brings the image's resolution to $390 \times 390$ and $375 \times 375$ from $400 \times 400$. Table-\ref{table:rob}  demonstrate that, despite the significant distortion in the watermarked images, the recovered watermark is still intact, and the values for $SSIM$ are close to one. Table-\ref{table:rob}  depicts the accuracy of $D_1$ in extracting the watermark (perceptual hash) and successful bit-by-bit comparison with the computed perceptual hash for different content-preserving image manipulations.  Fig-\ref{watermark_att} depicts the value of computed perceptual hash after applying different content-preserving image manipulations.

%\textbf{Performance of decoders ($D_2$ and $D_3$}): In the second phase of watermark extraction, two decoders are used for content and source authentication. The residual image can be reconstructed using $D_2$, and the cryptographic hash can be extracted from the reconstructed residual image using $D_3$. These decoders do not need to be robust (no distortion added in the watermarked image while training) because the cryptographic hash changes for only a single pixel change in the watermarked image. Thus, $D_2$ fails to reconstruct the residual image correctly, subsequently failing $D_3$ as well. This is the reason no robustness analysis is carried out on the decoders $D_2$ and $D_3$. 

\begin{table}[!htb]
\small
\fontsize{6.8pt}{9.5pt}\selectfont
\centering
\caption{Accuracy of $D_1$ when different content-preserving image manipulations are performed on the image. }
\begin{tabular}{|c|c|c|c|}

\hline
\textbf{Attack}                                              & \textbf{CelebA (\%)} & \textbf{Mirflicker (\%)} & \textbf{Merged (\%)} \\ \hline
\textbf{JPEG compression (Q=10)}                            & 93              & 91.8               & 91.9               \\ \hline
\textit{\textbf{JPEG compression (Q=30)}}                   & 96               & 95.1                 & 95.8            \\ \hline
\textit{\textbf{JPEG compression (Q=50)}}                   & 100           & 100                  & 100            \\ \hline
\textit{\textbf{Gaussian Blur (3 $\times$ 3)}} & 100             & 100                  & 100             \\ \hline
\textit{\textbf{Gaussian Blur (5 $\times$ 5)}} & 100             & 100                  & 100              \\ \hline
\textbf{Gaussian Blur (7 $\times$ 7)}          & 100             & 100                 & 100             \\ \hline
\textbf{Salt and Pepper noise}                              & 99.2             & 98.5                & 98.7            \\ \hline
\textbf{Gaussian Noise}                                     & 100             & 100                 & 100             \\ \hline
\textbf{Poisson noise}                                      & 100             & 100                 & 100             \\ \hline
\textbf{Cropping (20 $\times$ 20)}               & 97.7            & 98.3                & 98,2            \\ \hline
\textbf{Cropping (40 $\times$ 40)}             & 94.2            & 97.3                & 96.5            \\ \hline
\textbf{Cropping (60 $\times$ 60)}             & 88.3            & 92.4                & 89,9            \\ \hline
\textbf{Edge Cropping (390 $\times$ 390)}      & 100             & 100                 & 100             \\ \hline
\textbf{Edge Cropping (375 $\times$ 375)}      & 98.4            & 99                  & 98.7            \\ \hline
\end{tabular}

\label{table:rob}
\end{table}

 \begin{table*}[!ht]
 \fontsize{7pt}{10.5pt}\selectfont
 \caption{Comparison of our technique with the best known dual watermarking techniques with success against different attacks and its image quality index}
\begin{tabular}{|c|c|c|c|c|c|c|c|c|c|c|}
\hline
                                                                          &                     & \cite{liu2016blind}      & \cite{kumar2020dual}     & \cite{9214433}          & \cite{dual-2}             & \cite{dual-3}             & \cite{dual-4}           & \cite{dual-5}       & \cite{dual-6}        & Ours  \\ \hline
\multirow{2}{*}{Application}                                              & Copyright           & Yes       & No        & No            & Yes            & Yes             & Yes              & No        & Yes       & Yes   \\ \cline{2-11} 
                                                                          & Authentication      & Yes       & Yes       & Yes           & No             & No              & Yes              & Yes       & Yes       & Yes   \\ \hline
                                                  Technique               &            & DWT \& LSB & DWT \& SVD & DCT \& Spatial & Spatial  & Spatial \& DCT. & Wavlet  & DCT \& DWT & DCT \& DWT & DNN   \\ \hline
\multirow{2}{*}{\begin{tabular}[c]{@{}c@{}}Image \\ Quality\end{tabular}} & PSNR                & 40.32     & 36.97     & 41.27         & 38.9           & 54              & 39               & 37        & 46.53     & 46.87 \\ \cline{2-11} 
                                                                          & SSIM                & 0.98      & 0.99      & 0.99          & 0.99           & Na              & Na               & Na        & Na        & 0.99  \\ \hline
\multirow{9}{*}{Attack}                                                   & Salt \& pepper  & Yes       & Yes       & Yes           & Yes            & Yes             & Yes              & Yes       & Yes       & Yes   \\ \cline{2-11} 
                                                                          & JPEG Comp    & Yes       & Yes       & Yes           & Yes            & No              & Yes              & Yes       & Yes       & Yes   \\ \cline{2-11} 
                                                                          & Gaussian filter     & Yes       & Yes       & Yes           & Yes            & No              & Yes              & Yes       & No        & Yes   \\ \cline{2-11} 
                                                                          & Resizing            & Yes       & No        & Yes           & No             & Yes             & No               & No        & Yes       & Yes   \\ \cline{2-11} 
                                                                          & Cropping            & Yes       & Yes       & Yes           & Yes            & Yes             & Yes              & Yes       & Yes       & Yes   \\ \cline{2-11} 
                                                                          & Blurring            & Yes       & No        & No            & No             & No              & No               & No        & No        & Yes   \\ \cline{2-11} 
                                                                          & Median filter    & No        & Yes       & Yes           & No             & Yes             & Yes              & Yes       & Yes       & Yes   \\ \cline{2-11} 
                                                                          & Overwriting         & No        & No        & No            & No             & No              & No               & No        & No        & Yes   \\ \cline{2-11} 
                                                                          & Surrogate Model     & No        & No        & No            & No             & No              & No               & No        & No        & Yes   \\ \hline
\end{tabular}

\label{tab:attack}
\end{table*}

\subsection{Robustness Against Overwriting}
In the case of an overwriting attack, the attacker trains a surrogate encoder to overwrite the watermark of a watermarked image such that the target decoder decodes the attacker's watermark instead of the original watermark. Thus, the attacker claims ownership of the input watermarked image by overwriting its watermark. In our case, the overwriting attack is performed on decoder $D_1$, which is used for digital content ownership verification. To show the robustness of our technique against overwriting attack, we have trained a $U$-$Net$ based encoder using the same embedding procedure used for embedding perceptual hash. This encoder takes $50k$ watermarked images generated from our technique on the $Mirflickr$ dataset (cover images) and randomly sampled $64$-bit binary values (watermarks similar to perceptual hash)  as inputs to produce the new watermarked image. When these new watermarked images are passed through decoder $D_1$, the accuracy is still $92\%$. This means $D_1$ is able to extract the original watermark even if the overwriting attack is performed. The reason for withstanding the overwriting attack is following two-stage watermarking: in the first stage, $E_1$ reconstructs $W_1$ while in the second stage, $E_2$ generates {\em R}, which is added to  $W_1$ to obtain $W_2$.  $D_1$ is trained in the pipeline using loss functions to decode the perceptual hash even in the presence of a cryptographic hash as the watermark. This makes  $D_1$ robust and capable of extracting the watermark (perceptual hash) embedded using our technique and withstand overwrite attack.

%$D_1$ fails to extract the embedded perceptual hash after adding a substantial amount of noise, but there are visible distortions in the watermarked image.%why clarify

\subsection{Robustness Against Surrogate Model Attack}
A surrogate encoder can be trained to create watermarked images where the $SHA$-$3$ hash of the image is used as the watermark. The aim of these watermarked images is to fail our decoders $D_2$ and $D_3$ such that our decoders successfully decode the watermark. This will compromise security as the unintended content is authenticated successfully. To tackle this
in our technique, the cryptographic hash is extracted from the watermarked image and further compared with the computed cryptographic hash, which cannot tolerate single-pixel change. Hence, $D_2$ fails to reconstruct the residual image correctly for any minute distortion in the watermarked image, leading to the failure of $D_3$ in extracting the embedded hash. We have used two decoders, $D_2$ and $D_3$, such that we are not directly decoding but rather first reconstructing and then decoding.  To show the robustness of our technique against the surrogate model attack, we have trained three $U$-$Net$ based encoders with different layers and three decoders based on spatial transformer using the same procedure used for embedding cryptographic hash. Training is done on $50k$ images from the $Mirflickr$ dataset (cover images) and its corresponding cryptographic hash (watermark) to produce the watermarked image. When these watermarked images are passed through decoders $D_2$ and $D_3$ for verification, the decoders fail for every case which implies that the decoders could identify the original encoder and surrogate encoder. The attackers must know the whole encoding and decoding procedure without changing even one pixel in the process to succeed, which is nearly impossible. In this way, the cryptographic hash serves the purpose of content and source authentication by eliminating the possibility of surrogate model attacks.

\section{ Comparison with Previous Methods} \label{comp}
%Qualitative comparisons are essential to highlight the effectiveness of the proposed method. Incorporating visual illustrations and results will enhance the paper's overall impact.

%As already mentioned, our technique is the first deep learning-based dual watermarking technique. In order to show the quality and effectiveness of the proposed technique,
In this section, we compare our technique with the best-known results in the domain of traditional dual-image watermarking techniques. As shown in Table-\ref{tab:attack}, the proposed technique not only has a broader scope of application but is also robust against a broader attack surface while having a high $PSNR$ and $SSIM$ value among existing dual watermarking techniques. The robustness of different dual watermarking techniques against various forms of content-preserving image manipulation is shown in Table-\ref{tab:attack}. Our technique is robust against all attacks for protecting digital content copyright, as discussed in dual image watermarking techniques. Significant advantages are offered in terms of preventing overwriting and surrogate model attacks and facilitating source authentication for watermarked images, which is not available in existing techniques. It possesses the capability to identify even the slightest single-pixel alterations in received images. In summary, our proposed technique goes beyond traditional approaches by addressing the shortcomings. Our technique achieves a high level of visual quality as compared to existing deep learning-based single watermarking techniques~\cite{ref4,deepwatermark1,lacuna4}~\footnote{Among the three work \cite{ref4} reported $PSNR$ of $41.2dB$ and $SSIM$ of $0.98$,~\cite{deepwatermark1} reported $PSNR$ of $39.72dB$ and~\cite{lacuna4} reported $PSNR$ of $41.81dB$ and $SSIM$ of $0.99$.  }. This ensures that watermarked images maintain a high level of similarity with the cover image.  In addition, we achieved a similarity index of 96\% and an accuracy of 95\% (the average accuracy of all three decoders).

\section{Conclusion and Future Work}
\label{Conclusion and Future Work}
 We proposed a deep learning-based dual watermarking technique for source and image authentication as well as protecting digital content copyright. The technique embeds perceptual hash for protecting digital content copyright and cryptographic hash for content and source authentication. By embedding two separate watermarks into the cover image, we achieve a higher level of security against various content-preserving image manipulations while maintaining good perceptual quality. We demonstrated the system's ability to extract both watermarks accurately, ensuring high image security and eliminating the possibility of watermark overwriting and surrogate model attacks.  We carefully designed the loss functions and training strategy. Comprehensive experiments show that our technique is robust and doesn't affect the image's visual quality. We analyzed our model's efficiency in different scenarios, and the results show that our technique outperforms existing techniques. This illustrates our proposed model's high level of robustness and accuracy.

\bibliography{refereces} 
\bibliographystyle{ieeetr}

\end{document}